\documentclass{ws-procs9x6}

\usepackage{amsmath,epsfig,mathrsfs}
\usepackage[american]{babel}

\newcommand*{\dd}{\mathrm d}

\newcommand*{\halb}{\frac{1}{2}}

\newcommand{\DSGl}{{Dyson}--{Schwinger} equation}
\newcommand{\Tr}{\mathrm{Tr}\,}
\newcommand{\intsum}[1]{\ensuremath{   \int_{#1} \! \! \! \! \! \! \Sigma}}

\begin{document}

\title{Scalar O(N) Model at Finite Temperature --- 2PI~Effective Potential
  in Different Approximations}

\author{J\"urgen Baacke \lowercase{and} \underline{Stefan Michalski}}

\address{  Universit\"at Dortmund \\
  Institut f\"ur Physik,  Otto-Hahn-Str. 4 \\
  D--44221 Dortmund, Germany}
\maketitle

\abstracts{We calculate the two-particle irreducible (2PI) 
  effective potential of the $O(N)$ linear sigma model in 1+1 dimensions.
  The approximations we use are the next-to-leading order of a
  $1/N$ expansion (for arbitrary $N$)
  and a kind of ``resummed loop approximation'' for $N=1$.
  We show that the effective potential of the $1/N$ expansion is convex for
  $N=4$ and $N=10$ whereas it is not for the ``loop'' expansion
  and the case $N=1$ of the $1/N$ expansion.
}

\section{Introduction}
Inspired by earlier analyses out of equilibrium
and in thermal equilibrium~\cite{earlier} 
 we are interested in the phase structure
of the $O(N)$ linear sigma model in different approximations.
Since renormalization is quite a task in 3+1 dimensions we first
carry out this analysis in 1+1 dimensions. 
For all details the reader is referred to a more comprehensive 
analysis published recently~\cite{Baacke:2004dp}.

\section{Basic Equations}
\subsection{Classical action}
We take the following classical action for the $O(N)$ linear sigma model 
with spontaneous symmetry breaking in 1+1 dimensions
\begin{equation}
  \label{eq:lagrange}
  \mathcal{S}[\Phi]=\int \dd^2x\ 
  \frac{1}{2} (\partial_\mu \vec\Phi)^2 - \frac{\lambda}{4N}
  \left( \vec\Phi^2 -N v^2 \right)^2
  \quad\textrm{where}\quad
  \vec\Phi=(\Phi_1,\dots,\Phi_N)\ .
\end{equation}

In order to consider finite temperature we use the \textnormal{Matsubara}
formalism where the \textnormal{Minkowski}an integral over momentum space 
is transformed to a sum and an integral over spatial momenta.
We will use the short-hand notation
$ T \sum_{\omega_n} \int\frac{\rm dp}{2\pi} \equiv \intsum{p}$ .
For the effective action we consider a homogeneous background condensate
$\vec\phi = \langle \vec\Phi \rangle = (\phi,\phi,\dots,\phi)$ which
can be $O(N)$-rotated such that it points only in the 1-direction
$\vec\phi = \left( \sqrt{N} \phi, 0, \dots, 0 \right)$. 

\subsection{2PI effective action}
The two-particle irreducible (2PI) effective action~\cite{CJT} reads
\begin{equation}
  \label{eq:effective action}
  \Gamma[\phi,G] = \mathcal{S}\left[\sqrt{N}\phi\right] 
  + \halb \Tr\ (i\mathcal{D}^{-1}G-1)
  + \frac{i}{2} \ln\det G^{-1}D_0 + \Gamma_2[G,\phi] \ .
\end{equation}
Here the two-point function $G$ is a $N\times N$ matrix which is
$O(N)$ symmetric,
$  G = \mathrm{diag}\left[ G_\sigma, G_\pi, \dots, G_\pi \right] $,
and $\mathcal{D}$ denotes the analogous matrix of classical Green functions 
with
$
  i \mathcal{D}_{\sigma,\pi}^{-1}(k) = k^2 - \lambda(f_{\sigma,\pi}\
  \phi^2-v^2)$
and $f_{\sigma,\pi}=3,1$.
The explicit form of all higher-order corrections 
denoted by $\Gamma_2[G,\phi]$ is related to the type of approximation
used.

\subsubsection{Next-to-leading order of $1/N$ expansion}
The $1/N$ expansion is a systematic expansion of the effective
action in powers of $1/N$. 
The classical and one-loop part of
the effective action is of leading order $\mathcal{O}(N)$
whereas all further contributions (except for the double-bubble graph)
are of higher order. Here we will only take into account the contributions
of next-to-leading order (NLO) of the 2PI-$1/N$ 
expansion\footnote{Due to resummation contributions of all powers of
  $1/N$ contribute as well. Though the order is determined from the
2PI graphs of the action.}
Therefore, higher loop contributions are separated into three parts
\[ \Gamma_2[\phi,G] = \Gamma_2^\mathrm{db}[G] + 
    \Gamma_2^\textrm{pearls}[G] +  \Gamma_2^\textrm{sunset}[\phi,G]
    \ .
\]
The first term (so called double-bubble) contributes both at leading 
and next-to-leading order of $1/N$ (cf. larger articles for 
details~\cite{Baacke:2004dp,Aarts:2002dj})
To sum up all contributions of next-to-leading order (NLO)
we use the functional~\cite{Aarts:2002dj}
\begin{equation}
  \label{eq:Gamma_2^pearls}
  \Gamma_2^\textrm{pearls}[G] = \frac{i}{2} \intsum{p}\ 
  \left\{ 
  \ln \left[1+ i \frac{\lambda}{N} \Tr \mathcal{F}(p) \right]
    - i \frac{\lambda}{N} \Tr\mathcal{F}(p) \right\}
    \ ,
\end{equation}
where $\Tr \mathcal{F}$ is the trace of all fish graphs
\begin{equation}
  \label{eq:fish}
  \Tr \mathcal{F}(p)= \mathcal{F}_\sigma(p) + (N-1) \mathcal{F}_\pi(p)
  \quad\textrm{with}\quad
  \mathcal{F}_*(p) = \intsum{p}\ G_*(k)\, G_*(k+p) \ .
\end{equation}
The subtracted graph in Eq.~(\ref{eq:Gamma_2^pearls}) is the NLO part
of the double-bubble  which is dealt with separately~\cite{Baacke:2004dp}.

The generalization of sunset diagrams in the effective action
is achieved by cutting a sigma line of $\Gamma_2^\textrm{pearls}$ and 
pinning the two open legs to the background by multiplying by a factor of
$N\phi^2$
\begin{equation}
  \label{eq:Gamma_2^sunset}
  \Gamma_2^\textrm{sunset}[\phi,G] =i \lambda\ \phi^2\ \intsum{p}\ 
  G_\sigma(p) \frac{ i \frac{\lambda}{N} \Tr \mathcal{F}(p)}
  {1 + i\frac{\lambda}{N} \Tr \mathcal{F}(p)}\ .
\end{equation}
Expanding this in powers of $\mathcal{F}$ one finds graphs of the
generalized sunset kind (see again our larger 
article~\cite{Baacke:2004dp} for figures and more details).

\subsubsection{``Loop'' expansion for $N=1$}
Since for $N=1$ the $1/N$ expansion is somewhat meaningless,
one could have the idea to improve a loop expansion by summing up 
all pearls and generalized sunset graphs as in 
Eqs.~\eqref{eq:Gamma_2^pearls} and~\eqref{eq:Gamma_2^sunset}. 
We will call this ``loop expansion''
although this is not literally correct. We take into account
the same graphs (topologically) as in the $1/N$ expansion
but with a combinatorical factor that disregards their order of $1/N$.
The respective expression for the resummation of pearls is 
(note the additional combinatorical factor)
\begin{equation}
  \label{eq:Gamma_2^pearlsloop}
  \Gamma_2^\textrm{pearls}[G] = \frac{i}{2} \intsum{p}\
  \left\{ 
  \ln \left[1+ 3i\lambda\mathcal{F}(p) \right]
    - i3\lambda \mathcal{F}(p) 
    + 3 \lambda^2 \left[  \mathcal{F}(p)\right]^2
  \right\}
    \ .
\end{equation}
The subtracted graphs in Eq.~(\ref{eq:Gamma_2^pearlsloop}) 
have to be dealt with separately due to different combinatorical factors.
The last term in  Eq.~(\ref{eq:Gamma_2^pearlsloop}) is the
basketball graph which has a different topology than the
graphs with more than two vertices.
The sum of generalized sunset graphs is obtained in an analogous way
to the $1/N$ expression \eqref{eq:Gamma_2^sunset}. It reads
\begin{equation}
  \label{eq:Gamma_2^sunsetloop}
  \Gamma_2^\textrm{sunset}[\phi,G] =i\lambda\ \phi^2\ \intsum{p}\ 
  G_\sigma(p) \frac{ 3i\lambda \mathcal{F}(p)}
  {1 + 3i\lambda\mathcal{F}(p)}
  + 6 \lambda^2 \mathcal{F}(p) G(p)
  \ .
\end{equation}

\subsection{Dyson--Schwinger Equation}
In order to calculate the 1PI effective potential from the 2PI effective
action we have to solve the \DSGl\ $\delta \Gamma / \delta G = 0$
for the 2-point function.
Using the convention 
$iG^{-1}_*(p) = i\mathcal{D}_*^{-1} - \Sigma_*(p)$,
where  $*=\sigma,\pi$,
we can express the \DSGl\ in terms of the self-energy~$\Sigma(p)$
\begin{subequations}
  \label{eq:Dyson-Schwinger}
  \begin{eqnarray}
    \Sigma_\sigma(p) &=& 3 \frac{\lambda}{N}\ \mathcal{B}_\sigma 
    + (N-1) \frac{\lambda}{N} \mathcal{B}_\pi
    - 2\, \frac{\delta}{\delta G_\sigma(p)}
    \left( \Gamma_2^\textrm{pearls} + \Gamma_2^\textrm{sunset} 
    \right) \\
    \Sigma_\pi(p) &=& \frac{\lambda}{N}\ \mathcal{B}_\sigma 
    + (N+1) \frac{\lambda}{N} \mathcal{B}_\pi
    - 2\, \frac{\delta}{\delta G_\pi(p)}
    \left( \Gamma_2^\textrm{pearls} + \Gamma_2^\textrm{sunset}
    \right)
    \ .
  \end{eqnarray}
\end{subequations}
The 1PI effective action is obtained by substituting a solution $G(\phi)$
of Eqs.~\eqref{eq:Dyson-Schwinger} into $\Gamma[\phi,G]$. We will plot
the 1PI effective potential that differs from that only by a 
total factor of volume times temperature.


\section{Numerical Results}
For a given temperature $T$ and different values of $\phi$
we numerically solve the \DSGl~\eqref{eq:Dyson-Schwinger} 
by iteration.

\subsection{Next-to-leading order of 2PI $1/N$ expansion}
We take here a value of $\lambda=0.5$ for the coupling constant
and show results for $N=1$, $N=4$ and $N=10$.
\begin{figure}
  \centering
  \epsfig{file=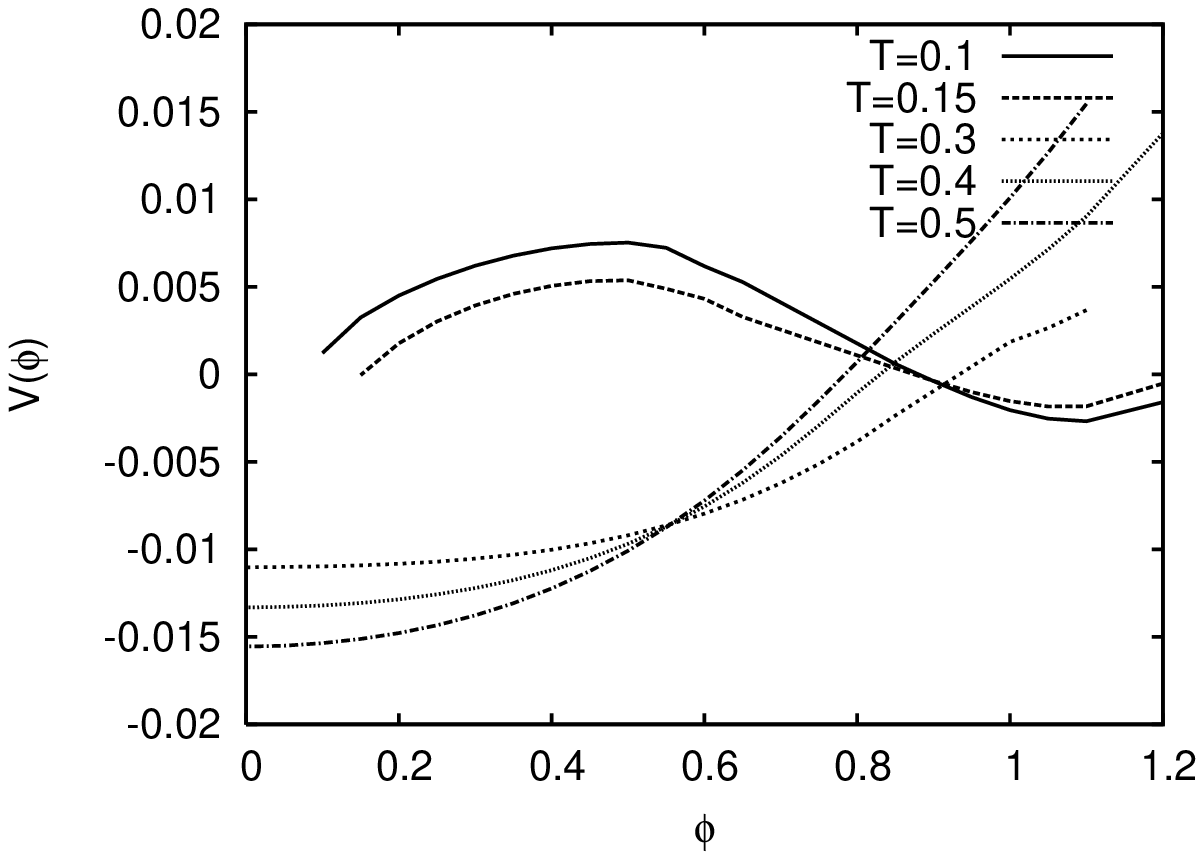,width=.3\columnwidth}
  \hfill
  \epsfig{file=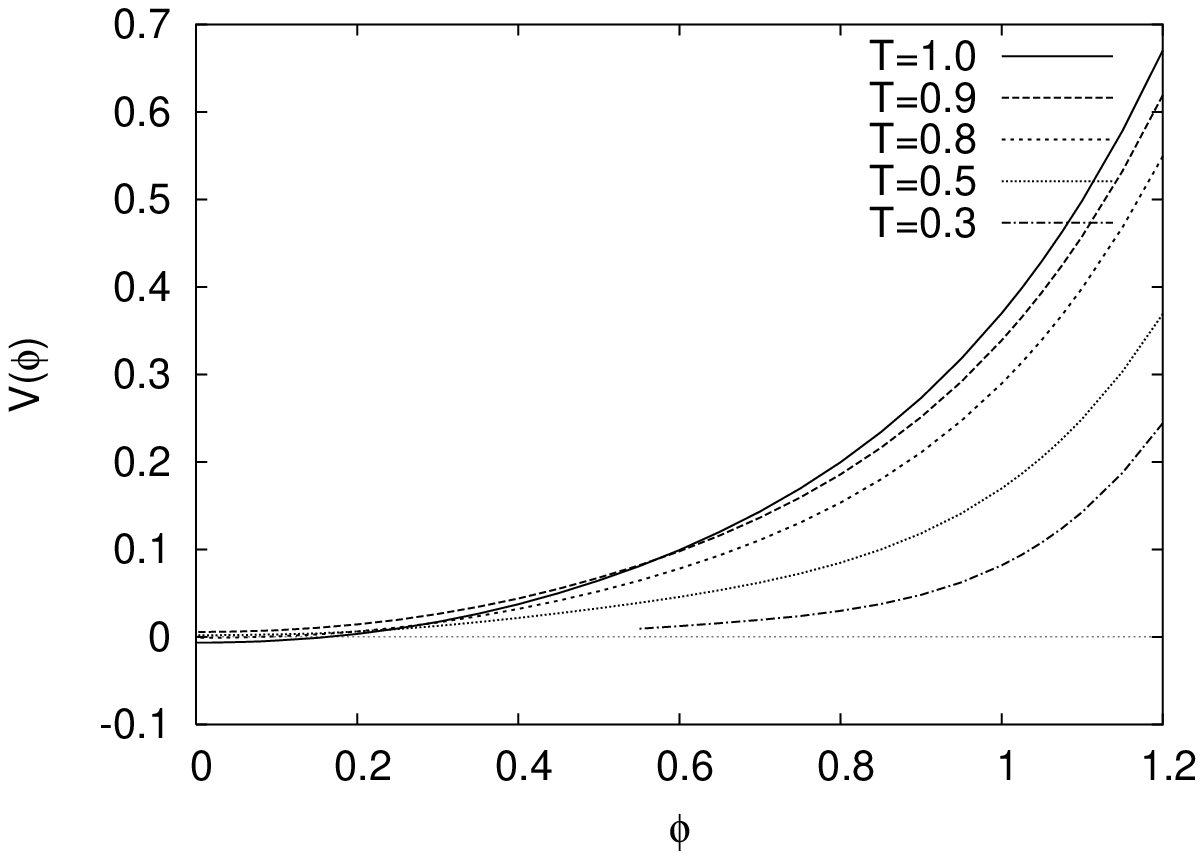,width=.3\columnwidth}
  \hfill
  \epsfig{file=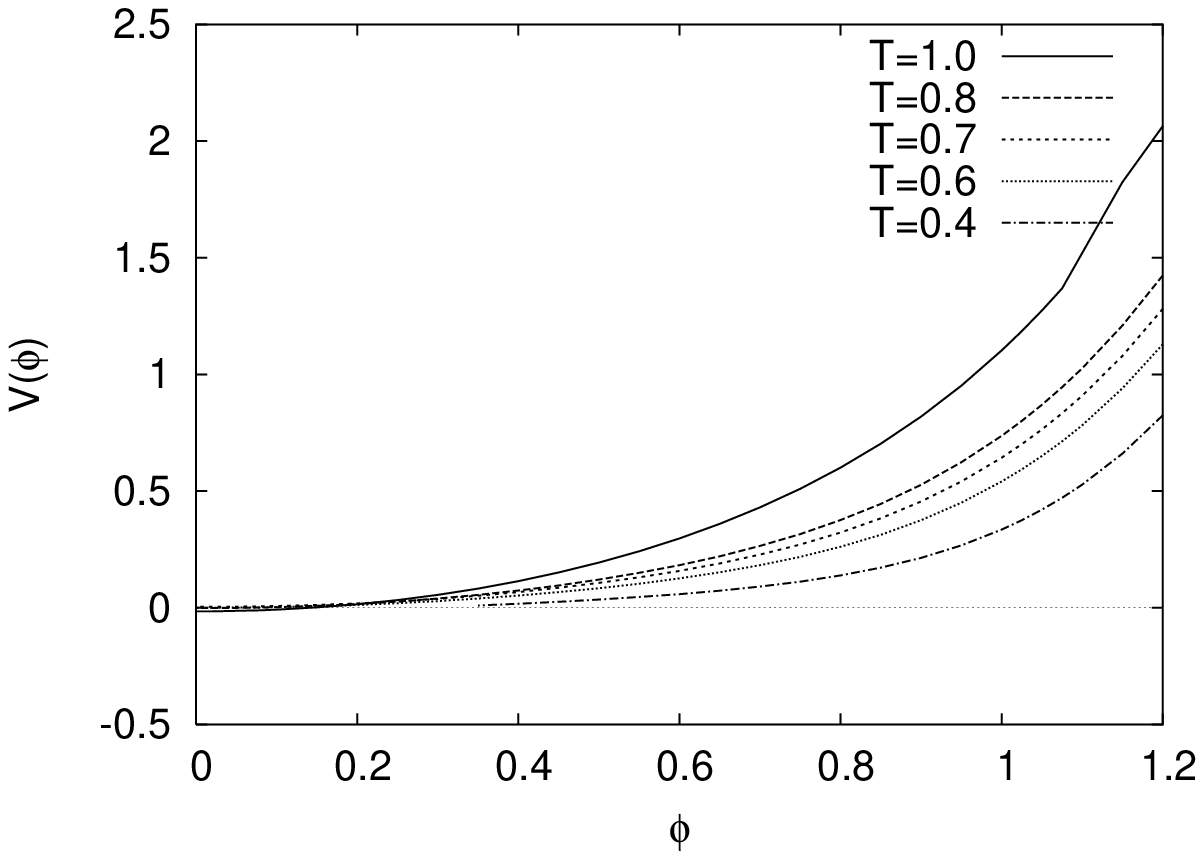,width=.3\columnwidth}

  \caption{Shape of the effective potential in the 2PI $1/N$ approximation
    at NLO. From left to right: $N=1$ and $\lambda=0.1$ (not normalized), 
    $N=4$ and $\lambda=0.5$, $N=10$ and $\lambda=0.5$.}
  \label{fig:1/N}
\end{figure}
Figure~\ref{fig:1/N} indicates that there is only one vacuum at $\phi=0$ 
for the considered temperatures and the cases $N=4$ and $N=10$.
The potential for $N=1$ shows signs of a false vacuum which is 
actually not expected in a $1/N$ expansion. Though for $N=1$ this
expansions is obviously not meaningful. For the sake of direct 
comparison we mention --- 
without showing a plot --- that for $N=1$ and $\lambda=0.5$ 
the $1/N$ potential is convex at temperatures $T=0.7$ and $T=0.8$, e.g.,
whereas it is not when using the ``loop expansion'' 
(see below and Fig.~\ref{fig:loop}).

\subsection{``Loop expansion'' for $N=1$}
\begin{figure}
  \centering
  \epsfig{file=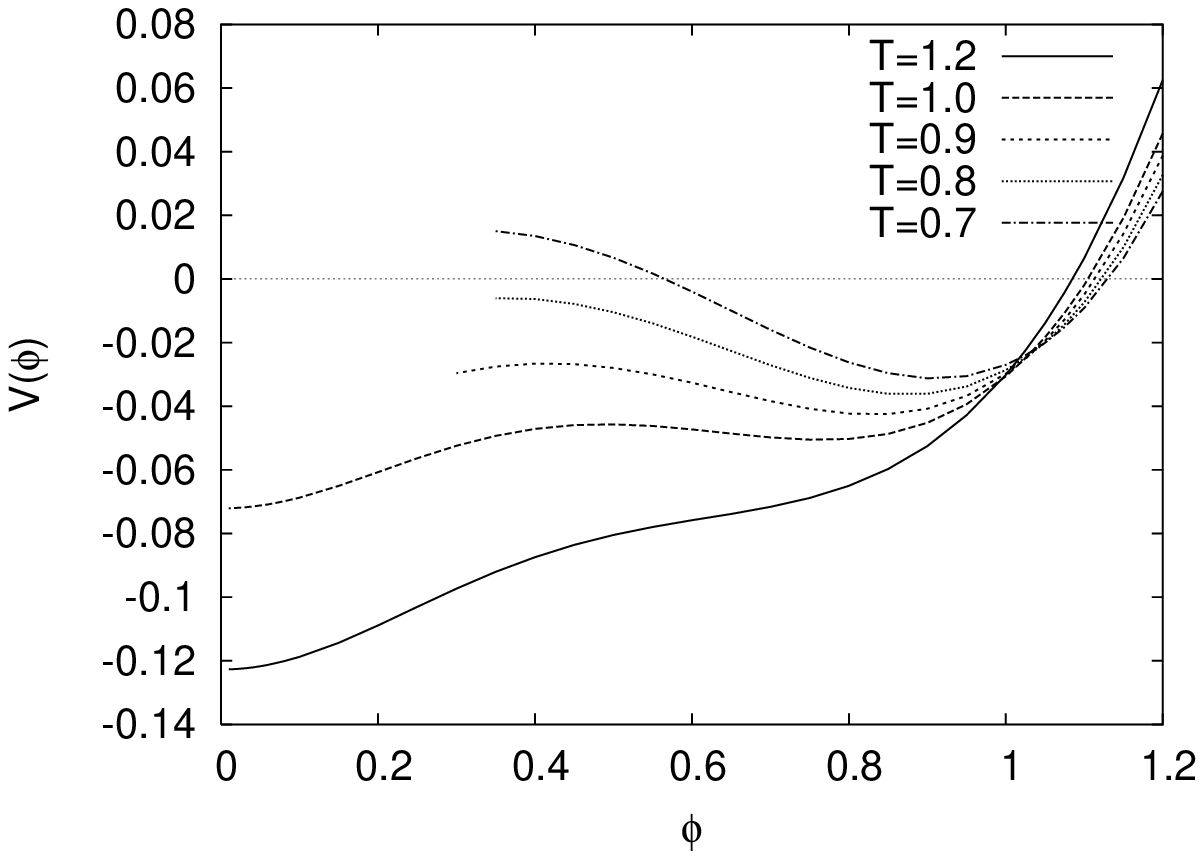,width=.3\columnwidth}
  \hfill
  \epsfig{file=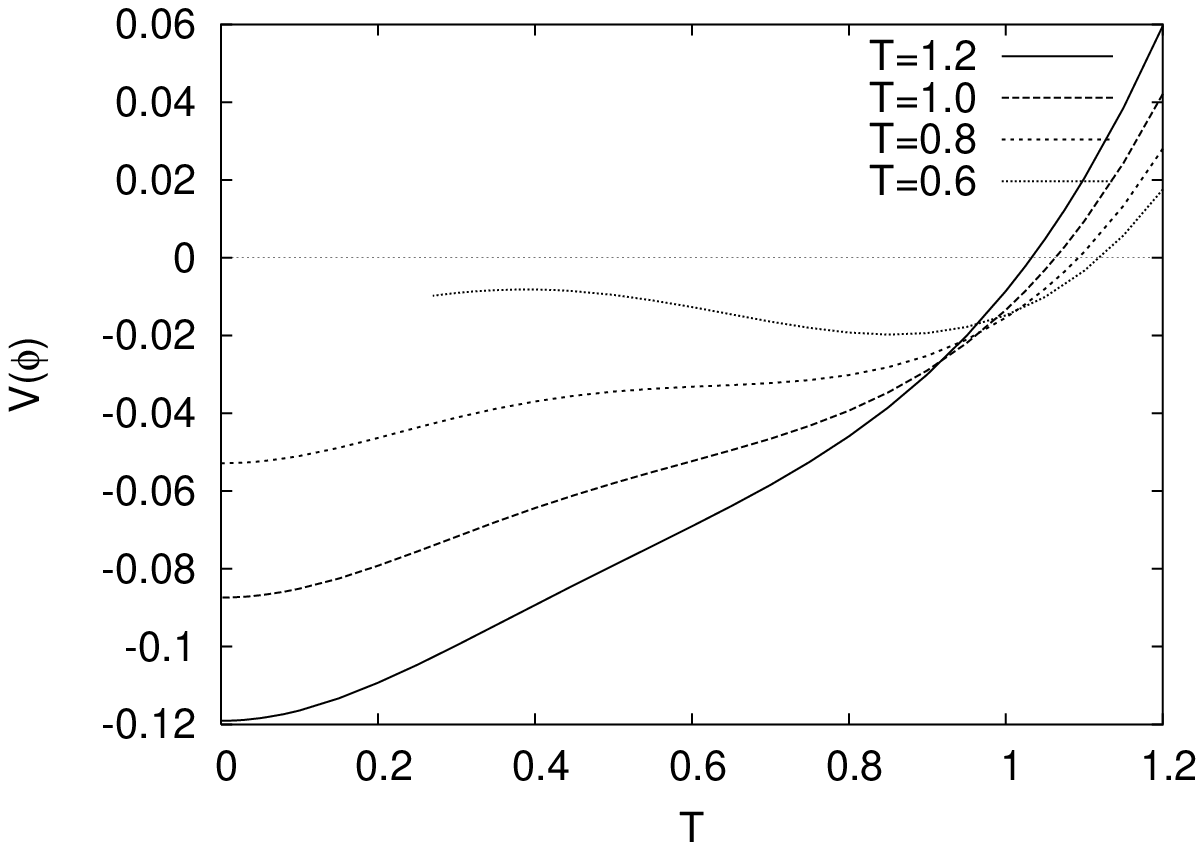,width=.3\columnwidth}
  \hfill
  \epsfig{file=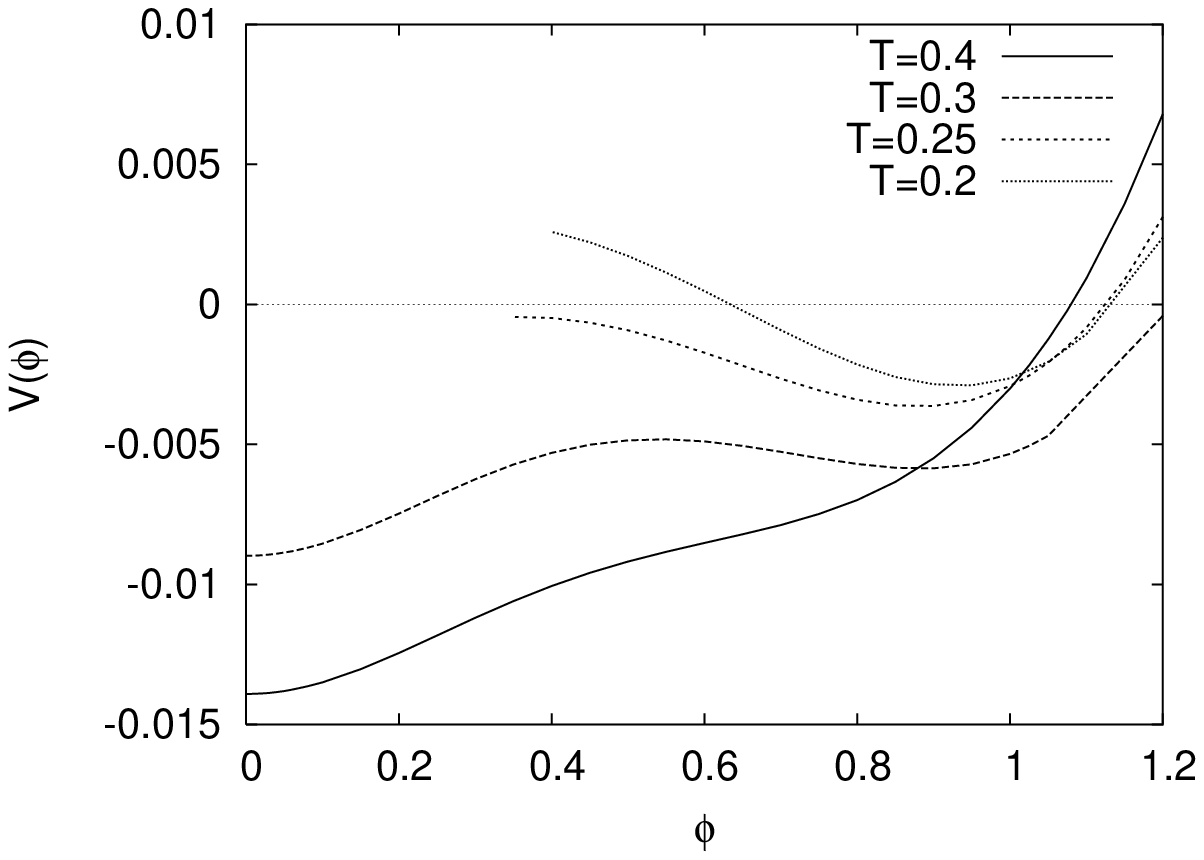,width=.3\columnwidth}
  
  \caption{Effective potential of the ``resummed loop expansion'' 
    at different temperatures for $\lambda=1$,
    $\lambda=0.5$ and $\lambda=0.1$.}
  \label{fig:loop}
\end{figure}

We display the effective potential in the resummed ``loop expansion'' 
(cf. Eqs.~\eqref{eq:Gamma_2^pearlsloop} and~\eqref{eq:Gamma_2^sunsetloop})
in Fig.~\ref{fig:loop} at different temperatures and for 
three values of the coupling constant. 
The potential
exhibits a typical structure with a false vacuum --- a clear sign of a 
first-order phase transition. For higher temperatures, e.g. $T=1.2$ and
$\lambda=1$ or $T=0.8$ and $\lambda=0.5$, the false vacuum has
disappeared but a ``relic'' consisting of two inflection points remains.


\section{Conclusion and Outlook}
We have solved the \DSGl\ to compute the effective potential of
the $O(N)$ linear sigma model in 1+1 dimensions both in a
``resummed loop'' expansion for $N=1$ and at NLO of a $1/N$ expansion
for arbitrary $N$.
For $N=4$ and $N=10$ the effective potential is convex for all parameters
we chose as expected from very old arguments~\cite{Coleman:1973ci}.

For $N=1$ we find (indications of) false vacua in both approximations.
The $1/N$ expansion seems to be meaningless for $N=1$ concerning the
shape of the effective potential. 
For the ``resummed loop expansion'' one has to admit that
this approximation only serves as an example of a non-systematic 
expansion and therefore the effective potential has a non-physical shape.
As stated above, further results can be found in a more comprehensive
publication\cite{Baacke:2004dp}.


\section*{Acknowledgments}
S.M. thanks all the organizers of \emph{SEWM 2004} for a wonderful
meeting in Helsinki. 
S.M. was supported by 
\emph{Deutsche Forschungsgemeinschaft} 
as a member of \emph{Graduiertenkolleg 841}.


\end{document}